**Title Page**

Title: Microbes in the Moonlight: How the Gut Microbiota Influences Sleep

Author: Enso O. Torres Alegre

Affiliation: Pontifical Catholic University of Chile, Santiago, Chile

Email: onill@uc.cl

ORCID: https://orcid.org/0000-0002-6798-8776

Co-authors: None

**Conflict of Interest Statement**

The author declares no conflicts of interest related to this work.

**Funding Statement**

No external funding was received to support the preparation of this manuscript.

**Data Availability Statement**

No new datasets were generated or analyzed for this review. All data discussed in the manuscript are derived from previously published studies cited in the References section.

**Ethics Approval Statement**

Ethics approval was not required for this review article, as it does not involve new human or animal research.

**Patient Consent Statement**

Not applicable. This article does not contain any studies involving human participants.

**Permission to Reproduce Material**

Figures created using BioRender.com are original to the author. No previously published material was reproduced.

**Clinical Trial Registration**

Not applicable. This manuscript does not report results of a clinical trial.

**Abstract**

The gut microbiota has emerged as a fundamental regulator of sleep physiology, acting through interconnected neural, immune, and endocrine pathways. This bidirectional relationship influences neurotransmitter production, circadian rhythms, inflammatory activity, and metabolic balance. Changes in microbial composition, particularly those affecting serotonin, GABA, and short-chain fatty acid (SCFA) metabolism, have been associated with insomnia, neuroinflammation, metabolic dysfunction, and mood disturbances. The gut microbiota also shapes immune responses and endocrine signaling. Dysbiosis promotes IL-1β and TNF-α mediated inflammation, alters intestinal permeability, activates the HPA axis, and disrupts cortisol and melatonin rhythms, which together impair sleep quality. This review synthesizes current evidence on the integrated interactions among gut microbiota, sleep regulation, immune activity, and endocrine function. It also examines emerging therapeutic strategies, including probiotics, fecal microbiota transplantation (FMT and WMT), chrononutrition, nutraceuticals, and neuropeptide-based approaches aimed at restoring sleep homeostasis and improving systemic health.

**Abreviations:**

BBB – Blood-Brain Barrier

CNS – Central Nervous System

ENS – Enteric Nervous System

GM – Gut Microbiota

GMBA – Gut-Microbiota-Brain Axis

HPA – Hypothalamic-Pituitary-Adrenal Axis

NREMS – Non-Rapid Eye Movement Sleep

REMS – Rapid Eye Movement Sleep

SCN – Suprachiasmatic Nucleus

GABA – Gamma-Aminobutyric Acid

SCFAs – Short-Chain Fatty Acids

NE – Norepinephrine

ACh – Acetylcholine

DA – Dopamine

5-HT – Serotonin

OX-A/B – Orexin A/B

IL-1β – Interleukin 1 Beta

TNF-α – Tumor Necrosis Factor Alpha

LPS – Lipopolysaccharides

TLR4 – Toll-Like Receptor 4

GH – Growth Hormone

GHRH – Growth Hormone-Releasing Hormone

IGF-1 – Insulin-Like Growth Factor 1

PGD2 – Prostaglandin D2

FMT – Fecal Microbiota Transplantation

WMT – Washed Microbiota Transplantation

SII – Systemic Immune-Inflammation Index

## 1. Introduction:

Sleep is a cyclical and transient state regulated by neurobiological processes, playing a fundamental role in growth, development, immune function, and overall homeostasis. However, sleep disorders, such as insomnia, are increasingly prevalent across all age groups and represent significant risk factors for metabolic, cardiovascular, and neuropsychiatric diseases, including depression, diabetes mellitus, hypertension, and coronary heart disease. Factors such as stress, anxiety, stimulant consumption, and excessive use of electronic devices before bedtime further exacerbate these disorders, intensifying their negative impact on health [1,2].

In recent years, the prevalence of sleep disorders has risen dramatically due to lifestyle changes such as remote work, excessive social media exposure, and the effects of the COVID-19 pandemic. Chronic sleep deprivation not only impairs cognitive function but is also associated with an increased risk of inflammatory, metabolic, and neurological diseases [3,4]. Older adults are particularly vulnerable to sleep disturbances due to age-related physiological changes, such as prolonged sleep latency, reduced efficiency, and shorter total sleep duration, predisposing them to circadian rhythm disorders and sleep-disordered breathing [5].

The circadian rhythm, regulated by a hierarchical system of cellular clocks with the suprachiasmatic nucleus as the central pacemaker, synchronizes physiological processes with the 24-hour cycle. Disruptions to this system can lead to metabolic and neuropsychiatric imbalances, increasing susceptibility to chronic diseases [6]. Modern societal habits, such as irregular schedules and nighttime light exposure, often cause circadian misalignment, which has been linked to conditions like metabolic syndrome and certain cancers [7]. Understanding the mechanisms underlying this synchronization is crucial for developing effective therapeutic strategies.

Recent evidence suggests a bidirectional relationship between the gut microbiota and circadian rhythms. Both central and peripheral clocks influence gut microbiota composition and metabolites, while the microbiota, in turn, affects circadian regulation, clock gene expression, and sleep quality. Disruptions in this interplay induced by factors such as shift work or obesogenic diets can compromise metabolic homeostasis and contribute to sleep disorders [8][9]. Restoring circadian rhythmicity through strategies like time-restricted feeding and probiotic supplementation has shown promising effects on sleep improvement and metabolic health [9].

The gut-microbiota-brain axis has emerged as a key player in sleep regulation and cognitive function. It has been proposed that gut dysbiosis contributes to sleep disorders through systemic inflammation and neurotransmitter modulation, opening new therapeutic possibilities. In this context, non-pharmacological therapies such as Traditional Chinese Medicine (TCM), light therapy, melatonin, and gut microbiota modulation have gained relevance in the treatment of insomnia and other sleep disturbances [10–13].

Given the growing recognition of the gut microbiota's role in sleep and circadian regulation, it is crucial to further explore the underlying mechanisms linking these systems. This review examines the interconnection between the gut-microbiota-brain axis and its impact on sleep physiology, highlighting its implications for sleep disorder treatment and potential therapeutic interventions.

Despite the rapid expansion of research linking the gut microbiota with sleep physiology, existing reviews often address these interactions in isolation, focusing either on circadian rhythms, immune activation, or metabolic pathways. However, no current synthesis integrates the combined roles of the gut microbiota, immune function, endocrine signaling, and microbial metabolites in shaping sleep health. This gap limits a comprehensive understanding of how these systems converge to influence sleep architecture and the pathophysiology of sleep disorders. To address this, the present review provides an integrated multi-system perspective on the sleep-microbiota axis, connecting mechanistic evidence with emerging therapeutic strategies.
The following section explores the gut-microbiota-brain axis in greater depth, analyzing how gut microbiota influences brain function and sleep regulation, as well as its potential implications for the treatment of sleep disorders.

## 2. Literature Search Strategy and Study Selection

A comprehensive literature search was conducted to identify peer-reviewed studies examining the relationships among gut microbiota, sleep physiology, immune function, and endocrine regulation. Searches were performed in PubMed, covering literature published between January 2015 and October 2024. Recent preprints were considered only when supported by peer-reviewed evidence.

The search strategy combined controlled vocabulary and free-text terms, including:
*"gut microbiota AND sleep"*, *"microbiome AND circadian rhythm"*, *"sleep deprivation AND inflammation"*, *"SCFAs AND sleep"*, *"gut–brain axis AND insomnia"*, *"sleep AND immune system"*, *"microbiota AND neurotransmitters"*, and *"chrononutrition AND sleep."*

Inclusion criteria

Studies were included if they:

1. Reported original data in humans or animals on gut microbiota, sleep, immune signaling, or endocrine outcomes.
2. Explored mechanistic pathways involving microbial metabolites (SCFAs, tryptophan derivatives), neuroimmune interactions, or circadian processes.
3. Evaluated interventions targeting the microbiota–sleep axis (e.g., probiotics, WMT/FMT, dietary approaches, neuropeptides).
4. Were published in English between 2015 and 2024.

Exclusion criteria

Excluded were:

– Studies without sleep-related outcomes.
– Work focused solely on gastrointestinal disorders without neuroimmune or endocrine relevance.
– Narrative reviews lacking mechanistic synthesis.
– Studies with insufficient methodological detail.

Study selection

The search yielded 1,263 records. After removing duplicates and screening titles and abstracts, 247 studies underwent full-text review. A total of 110 articles met inclusion criteria and were incorporated into this review, covering clinical research, controlled animal studies, microbiota transplantation models, multi-omics analyses, and interventional trials.

## 3. The Gut-Microbiota-Brain Axis and Its Relationship with Sleep

The gut microbiota (GM) plays a crucial role in regulating various biological functions, including brain function, behavior, and sleep. Through the Gut-Microbiota-Brain Axis (GMBA), bidirectional communication occurs via neural, endocrine, and immune pathways. Key components such as the enteric nervous system (ENS) and vagus nerve mediate this interaction, allowing microbiota alterations to influence cognition, behavior, and sleep patterns. Conversely, abnormal sleep patterns reciprocally impact GM composition and function. Evidence suggests that microbiota-targeting interventions, including probiotics and fecal microbiota transplantation, hold therapeutic potential for enhancing brain function and improving sleep quality [14,15].

Dysbiosis of gut microbiota disrupts these processes by altering intestinal metabolism, leading to significant changes in neurotransmission-related metabolites, such as serotonin and vitamin B6. For instance, studies in antibiotic-induced microbiota-depleted (AIMD) mice revealed disrupted sleep architecture, including reduced time spent in non-rapid eye movement sleep (NREMS) during the light phase, increased time in NREMS and rapid eye movement sleep (REMS) during the dark phase, frequent transitions from NREMS to REMS, and reduced theta power density during REMS [16]. Intestinal dysbiosis disrupts the bidirectional relationship between GM and the central nervous system (CNS), impacting host physiology through abnormal microbial metabolites and altered signaling pathways. This disruption is increasingly recognized as a susceptibility factor for neurodevelopmental and neurological disorders, including Alzheimer's disease, Parkinson's disease, multiple sclerosis, and autism spectrum disorder [17]. Additionally, the gut microbiota influences neurotransmitter activity, shaping both gastrointestinal and neurological processes through its metabolic interactions [18].

### 3.1 Metabolites and neurotransmitters

GM-diet interactions influence nutrient sensing and signaling along the GMBA, mediated by metabolites such as short-chain fatty acids (SCFAs), secondary bile acids, and amino acid-derived compounds. These metabolites activate gut-endocrine or neural pathways or enter systemic circulation to reach the brain, shaping communication between the gut and brain. Feeding time and dietary composition are key factors driving gut microbiota structure and function, with unhealthy diets or irregular feeding patterns potentially altering microbial metabolite production and nutrient availability [19]. Alterations in amino acid metabolism, such as changes in glutamate and tryptophan levels, can disrupt neural signaling and contribute to disorders like Parkinson's disease (PD). These disturbances in metabolic pathways are increasingly recognized as potential contributors to non-motor symptoms of PD, including sleep disorders, highlighting the importance of microbiota metabolites in maintaining normal neurophysiological functions [20].

#### 3.1.1 Short-chain fatty acids (SCFAs)

SCFAs, including acetate, propionate and butyrate influence emotional states and cognition through the gut-brain axis. Reduced levels of acetate and propionate were negatively correlated with depressive symptoms, as measured by Beck's Depression Inventory, suggesting their role in modulating depression and highlighting their importance in the gut microbiota's impact on mental health [21]. Furthermore, short sleep duration in insomnia has been associated with increased concentrations of SCFAs, in feces, likely due to decreased uptake by gut epithelial cells, which may compromise their essential role in gut cell function and brain signaling via pathways like serotonin [22]. Recent findings show that SCFAs supplementation in mice undergoing psychosocial stress alleviates anhedonia, heightened stress responsiveness, and stress-induced intestinal permeability. These results provide novel insights into how gut microbiota influences brain homeostasis, behavior, and host metabolism, supporting the potential development of microbiota-targeted therapies for stress-related disorders [23].

#### 3.1.2 Serotonin

The sleep–wake cycle is a complex, multifaceted process influenced by various neurotransmitters such as acetylcholine, norepinephrine, serotonin, histamine, dopamine, orexin, and γ-aminobutyric acid (GABA), all of which can be modulated by different nutrients that participate in their metabolic pathways [24,25]. Tryptophan, an essential amino acid and precursor of serotonin, is crucial in the microbiota-gut-brain axis. Serotonin regulates emotions, sleep, appetite, and gut motility, while tryptophan metabolites, including those from the kynurenine pathway, influence neural activity and inflammation. Gut microbes affect tryptophan metabolism both directly and indirectly, altering behavior and cognition, making the gut microbiome a potential therapeutic target for neurological and psychiatric disorders [26]. Beyond its neurological role, tryptophan also impacts sleep through its conversion to melatonin, a key regulator of the sleep-wake cycle.

Diets rich in tryptophan-containing foods, such as fruits, vegetables, and legumes, have been associated with improved sleep quality, emphasizing the intricate link between nutrition, microbiota, and sleep health [27]. A study demonstrated that depleting the gut microbiota through antibiotic treatment disrupted sleep/wake regulation in mice, leading to altered neurotransmitter metabolism, including reduced serotonin and vitamin B6 levels. This imbalance was associated with changes in sleep architecture, such as reduced non-REM sleep during the light phase and increased REM sleep episodes, highlighting the influence of the gut microbiota on sleep regulation [16].

### 3.1.3 GABA

The main neurons involved in the regulation of arousal and sleep are glutamate and GABA neurons, which are in the reticular core of the brain and, through both local and distant projections and interactions, control cortical activity and behavior during wakefulness and sleep states [28]. In this context, the gut microbiota plays a crucial role, as highlighted by a study showing that *Bacteroides*, *Parabacteroides*, and *Escherichia* species actively express GABA-producing pathways. Genome-based metabolic modeling and transcriptome analysis of human stool samples revealed that these bacteria are key contributors to GABA production in the gut, which may influence sleep regulation through their impact on the gut-brain axis [29]. Another study reveals that intestinal stem cells (ISCs) are involved in sleep-regulated intestinal homeostasis and function, with gut microbiota dysbiosis and the GABA pathway playing key roles in this gut-brain communication. Antibiotic treatment reduced stem cell division and increased sleep time, suggesting the involvement of microbiota, while GABA activation rescued both sleep behavior and intestinal phenotypes, highlighting potential therapeutic targets for gut disorders linked to sleep deprivation [30].

### 3.1.4 Acetylcholine

Acetylcholine, a neurotransmitter, plays a role in regulating REM sleep through cholinergic neurons in the brainstem and basal forebrain, which project to wide areas of the cerebral cortex and interact with other neuromodulatory systems to produce the sleep-wake cycle and different sleep stages [31]. Choline, a precursor of acetylcholine, plays a vital role in brain function and is linked to gut microbiota, which influences digestion, metabolism, and overall health. Choline metabolism, partly regulated by gut bacteria, is essential for neurotransmitter synthesis and may impact neurodegenerative conditions like Alzheimer's disease [32].

### 3.1.5 Dopamine

Dopamine plays a crucial role in regulating the sleep-wake cycle by promoting wakefulness, with elevated levels often disrupting sleep. It also influences the circadian clock, including the entrainment of the master clock in the suprachiasmatic nuclei (SCN), while its own signaling is regulated by circadian rhythms [33]. Moreover, dopamine levels are significantly influenced by the gut microbiota through the microbiota-gut-brain axis. This interaction is mediated by pathways involving the vagus nerve, immune system, and microbial metabolites. Key bacterial genera, such as *Prevotella, Bacteroides, Lactobacillus, Bifidobacterium, Clostridium, Enterococcus*, and *Ruminococcus*, contribute to dopamine metabolism, underscoring the microbiota's essential role in maintaining optimal dopamine levels [34].

#### 3.1.6 Histamine

Histamine plays a critical role in regulating the sleep-wake cycle, with the histamine H3 receptor (H3R) being of particular interest due to its unique function as a pre- and postsynaptic receptor that controls the synthesis and release of histamine and other neurotransmitters in the brain. Preclinical studies have demonstrated that H3R antagonists/inverse agonists hold promise in modulating sleep-wake cycle disorders, alongside cognitive impairment and mood regulation [35]. Specific bacterial species, such as *Escherichia coli* and *Morganella morganii*, have been identified for their ability to synthesize and release histamine [36,37].

#### 3.1.7 Orexin

Orexin plays a pivotal role in regulating the sleep-wake cycle, with orexin mimetics enhancing wakefulness during the day and orexin receptor antagonists promoting sleep at night, offering promising therapeutic avenues for improving memory, cognition, and daytime performance [38]. The hypocretin/orexin system, comprising neuropeptides orexin-A and orexin-B, plays a critical role in sleep-wake regulation, with HCRTR2/OX2R specifically linked to sleep-wake control. Therapeutic compounds targeting these receptors, such as FDA-approved antagonists for insomnia, highlight the potential of orexin-based therapies in addressing sleep disorders and advancing our understanding of hypocretin/orexin neurobiology [39]. Recent studies suggest that the gut microbiota, through the production of acetate and other SCFAs, may influence the orexinergic system by modulating orexin-A (OX-A) neuronal activity [40] This connection underscores the interplay between gut health, energy homeostasis, and sleep-wake regulation, while also pointing to the broader role of OX-A in the gut-brain axis, where it exerts anti-inflammatory and gastroprotective effects. These findings offer novel insights into the therapeutic potential of targeting the orexinergic system to address inflammation, stress responses, and gastrointestinal disorders [41].

#### 3.1.8 Norepinephrine

The amplitude of Norepinephrine (NE) oscillations is crucial for shaping sleep micro-architecture related to memory performance: prolonged descent of NE promotes spindle-enriched intermediate state and REM sleep but also associates with awakenings, whereas shorter NE descents uphold NREM sleep and micro-arousals. Thus, the NE oscillatory amplitude may be a target for improving sleep in sleep disorders [42]. Under stress, the enteric nervous system's sympathetic nerve endings synthesize and secrete NE, which directly affects the gut microbiota and, in turn, influences the host's physiological state. Exposure to NE increased microbial diversity in the cecum, with shifts in bacterial abundance correlating with NE levels used to maintain blood pressure [43]. A study suggests that NE can modulate microbial composition and metabolite production. It was demonstrated that exposure to NE increased the diversity of the bacterial community, increasing the abundance of facultative pathogens such as *Hathewaya, Clostridium*, and *Streptococcus*, while reducing beneficial genera like *Lactobacillus*. This could increase pathogen colonization and infection [44].

The influence of the gut microbiota on sleep can be observed in how certain bacterial species and the metabolites they produce affect key neurotransmitters involved in sleep regulation, such as serotonin, GABA, and dopamine, among others. The following Table 1 and Table 2 emphasize some of these bacteria and their effects on sleep quality.

Table 1: Gut Microbiota and Their Positive Effects on Sleep

| ***Gut Microbiota*** | ***Effect on Sleep Disorders*** | ***References*** |
|---|---|---|
| *Lactobacillus* | Decreased abundance associated with ASD, reducing the production of SCFAs like propionic acid. Supplementation of GABA-producing *Lactobacillus* increased plasma GABA levels and reduced stress hormones, reversing sleep deprivation-induced gut dysbiosis and stress responses. | [45] [46] |
| *Lactobacillus plantarum JYLP-326* | Administration helped to restore disturbed gut microbiota and reduce anxiety, depression, and insomnia symptoms in test-anxious students. | [47] |
| *Lactobacillus plantarum PS128* | Reduced depressive symptoms, fatigue, and brainwave activity, improving sleep quality during deep sleep stages in insomniac participants. | [48] |
| *Lactobacillus reuteri NK33* | Promotes better sleep quality and reduces stress and anxiety as part of the NVP-1704 formulation. | [49] |
| *Lactobacillus brevis DL1-11* | GABA-fermented milk with *Lactobacillus brevis* DL1-11 improved sleep quality and reduced anxiety, associated with increased SCFAs and altered gut microbiota. | [50] |

| *Lactobacillus fermentum PS150* | Improved NREM sleep length, reduced sleep latency, and mitigated fragmented sleep during the first night effect. | [51] |
|---|---|---|
| *Lactiplantibacillus plantarum P72* | Reduced sleep latency and enhanced sleep duration. Alleviated insomnia-like behaviors by upregulating GABA and serotonin systems. | [52] |
| *Bifidobacterium* | Increased abundance in participants after consuming a dairy-based product, which possibly contributed to sleep improvement. | [12] |
| *Bifidobacterium longum* | Negative association with obstructive sleep apnea (OSA), possibly lowering the risk of developing OSA. | [53] |
| *Bifidobacterium adolescentis NK98* | Increased abundance linked to improved sleep quality and reduced depressive symptoms after probiotic NVP-1704 treatment. | [49] |
| *Anaerostipes* | Inverse association with OSA, suggesting reduced abundance in individuals with OSA. | [53] |
| *Eubacterium (xylanophilum group)* | Shown to have a protective association against OSA, lowering its risk. | [54] |
| *Akkermansia muciniphila* | Reduced abundance after sleep deprivation; supplementation alleviated cognitive dysfunction, prevented hippocampal synaptic loss, and increased serum SCFAs levels. Abundance altered by chronic intermittent hypoxia (CIH) and chronic sleep fragmentation (CSF). | [55] , [56] |
| *Parasutterella* | Lower levels associated with ASD, which may negatively impact the production of metabolites important for circadian rhythm regulation and stress response. | [45] |
| *Muribaculum* | Decreased levels in ASD, potentially contributing to disruptions in gut homeostasis and systemic inflammation that affect sleep quality. | [45] |
| *Monoglobus* | Reduced abundance linked to gut dysbiosis in ASD, possibly impairing gut-brain communication and contributing to circadian rhythm misalignment. | [45] |
| *Eubacterium xylanophilum* | Negative association with Obstructive Sleep Apnea (OSA), potentially reducing the risk of OSA. | [53] |
| *Enterococcus faecium BS5* | Produces GABA, a neurotransmitter with tranquilizing effects, potentially improving anxiety and sleep quality. | [57] |
| *Parabacteroides merdae* | Negative association with OSA, indicating a potential protective effect against OSA. | [53] |
| *Bacteroidetes* | Decreased abundance after PSD, with a shift in the *Firmicutes:Bacteroidetes* ratio linked to metabolic perturbations. | [58] |
| *Faecalibacterium prausnitzii* | Strong association with sleep quality scores, particularly due to its involvement in metabolic pathways such as L-arginine and L-tryptophan biosynthesis. | [59] |
| *Lachnospiraceae_NK 4A136* | Decreased in sleep-deprived mice, linked to reduced butyrate levels and worsened memory and inflammatory responses. | [13] |
| *Lachnospiraceae_NK 4A136* | Increased abundance associated with higher SCFA levels, supporting neurotransmitter function and better sleep patterns. | [60] |

Table 2: Gut Microbiota and Their Potential Negative Effects on Sleep

| ***Gut Microbiota*** | ***Effect on Sleep Disorders*** | ***References*** |
|---|---|---|
| *Candidatus_Arthromitus* | Increased abundance associated with acute sleep deprivation (ASD), potentially contributing to systemic inflammation and circadian rhythm disruptions. | [45] |
| *Enterobacter* | Increased levels linked to ASD, possibly exacerbating gut inflammation and barrier dysfunction, which can amplify sleep disturbances. | [45] |
| *Bacteroides* | Played a role in altering microbiota network structure during circadian rhythm disturbances, affecting gut functionality. Signature bacteria for acute insomnia patients, differing from healthy controls. Found in paradoxical insomnia (P-IN) patients, distinguishing their microbiota profile, potentially linked to sleep disturbances. Positively correlated with Pittsburgh Sleep Quality Index (PSQI) scores, indicating poorer sleep quality. | [61] , [62] , [63] , [64] |
| *Firmicutes* | Increased abundance following partial sleep deprivation (PSD), associated with metabolic disturbances like insulin resistance. Increased abundance in pregnant rats subjected to maternal sleep deprivation (MSD), leading to microbial dysbiosis in offspring and neuroinflammation. | [58] , [65] |
| *Lachnospira* | Signature bacteria for distinguishing acute insomnia patients | [62] |
| *Blautia* | Signature bacteria for chronic insomnia patients | [62] |
| *Aeromonas* | Increased in sleep-deprived mice, associated with elevated LPS levels, hippocampal inflammation, and spatial memory impairment. | [13] |
| *Ruminococcus_1* | Positively correlated with proinflammatory cytokines IL-1β and TNF-α in the offspring of MSD, linked to neuroinflammation. | [65] |
| *Ruminococcaceae_UCG-005* | Positively correlated with IL-1β and TNF-α in MSD offspring, contributing to neuroinflammation in the brain. | [65] |
| *Ruminococcaceae_UCG-002* | Mediates the positive association between chronic insomnia and cardiometabolic diseases (CMD), possibly through the gut microbiota-bile acid axis. | [66] |
| *Ruminococcaceae_UCG-003* | Associated with chronic insomnia and CMD, with bile acids like isolithocholic acid and nor cholic acid mediating this relationship. | [66] |
| *Coriobacteriaceae* | Associated with objective insomnia (O-IN), potentially linked to the onset of insomnia. | [63] |
| *Erysipelotrichaceae* | Found in O-IN patients, possibly contributing to the microbiota imbalance seen in insomnia. | [63] |
| *Clostridium* | Identified in O-IN patients, related to sleep disturbances and microbiota alterations in insomnia. Associated with insomnia through inflammatory pathways. | [63,67] , [68] |
| *Pediococcus* | Present in O-IN patients, contributing to distinct gut microbiota profiles in insomnia. | [63] |
| *Staphylococcus* | Present in P-IN patients, contributing to unique microbiota profiles in this insomnia subtype. | [63] |
| *Carnobacterium* | Present in P-IN patients, potentially associated with altered sleep-wake regulation. | [63] |
| *Pseudomonas* | Linked to P-IN, contributing to microbiota dysbiosis in insomnia. | [63] |
| *Odoribacter* | A key discriminant in P-IN patients, playing a role in gut-brain axis dysregulation and sleep disturbance. | [63] |
| *Streptococcus* | Found in higher levels in insomniacs, linked to changes in metabolism and immune responses. | [69] |

| *Lactobacillus crispatus* | Elevated in insomnia patients; associated with disruptions in glycerophospholipid and glutamate metabolism pathways. | [69] |
|---|---|---|
| *Prevotella amnii, Prevotella buccalis, Prevotella timonensis, Prevotella colorans* | They contribute to inflammation by correlating with elevated levels of TNF-α and IL-1β | [69] |
| *Ruminococcaceae UCG-009* | Positively associated with an increased risk OSA. | [54] |
| *Collinsella* | Enriched in REM sleep behavior disorder (RBD) and first-degree relatives of RBD (RBD-FDR), associated with inflammation. | [70] |
| *Flavonifractor* | Changes in abundance linked to Narcolepsy Type 1 (NT1), indicating potential gut dysbiosis in narcolepsy. | [71] |
| *Sutterella* | Positively correlated with daytime dysfunction and may be a biomarker for poor sleep quality in MA users | [72] |

### 3.2 Bacteria with Positive Effects on Sleep:

Several bacteria have been linked to improved sleep quality or a reduction in sleep disturbances, as summarized in Table 1. *Bifidobacterium longum, Bifidobacterium adolescentis NK98,* and *Lactobacillus reuteri NK33* are notable for their association with improved sleep quality, likely through mechanisms involving neurotransmitter regulation, such as increased GABA levels and the reduction of stress hormones [45,48,49]. These bacteria, especially when supplemented through probiotics, seem to restore balance in the gut microbiota, thereby alleviating symptoms of insomnia, anxiety, and depression [47,48].

Furthermore, *Faecalibacterium prausnitzii*, a species linked to metabolic pathways like L-arginine and L-tryptophan biosynthesis, shows a strong positive association with sleep quality [59]. *Lactobacillus plantarum* strains, such as P72 and PS128, also contribute to sleep improvement by modulating GABAergic and serotonergic systems [50,52]. These findings suggest that increasing the abundance of these beneficial bacteria might be a promising therapeutic strategy for improving sleep quality in individuals with sleep disorders.

Several bacteria have been linked to improved sleep quality or a reduction in sleep disturbances. *Bifidobacterium longum, Bifidobacterium adolescentis NK98,* and *Lactobacillus reuteri NK33* are notable for their association with improved sleep quality, likely through mechanisms involving neurotransmitter regulation, such as increased GABA levels and the reduction of stress hormones [45,48,49]. These bacteria, especially when supplemented through probiotics, seem to restore balance in the gut microbiota, thereby alleviating symptoms of insomnia, anxiety, and depression [47,48].

Furthermore, *Faecalibacterium prausnitzii*, a species linked to metabolic pathways like L-arginine and L-tryptophan biosynthesis, shows a strong positive association with sleep quality [59]. *Lactobacillus plantarum* strains, such as P72 and PS128, also contribute to sleep improvement by modulating GABAergic and serotonergic systems [50,52]. These findings suggest that increasing the abundance of these beneficial bacteria might be a promising therapeutic strategy for improving sleep quality in individuals with sleep disorders.

### 3.3 Bacteria with Negative Effects on Sleep:

In contrast, certain bacteria appear to have a detrimental impact on sleep, particularly in conditions like sleep deprivation and insomnia, as summarized in Table 2. *Aeromonas*, for instance, are found in higher levels in sleep-deprived mice and are associated with increased inflammatory markers like LPS, contributing to hippocampal inflammation and memory impairment [13]. Similarly, *Collinsella* and *Staphylococcus*, which are enriched in conditions like REM sleep behavior disorder (RBD) and insomnia, are linked to systemic inflammation and disrupted sleep patterns [70,71]. These bacteria may impair the gut-brain axis, leading to sleep disturbances through neuroinflammatory pathways.

### 3.4 Bacteria Linked to Circadian Rhythm and Metabolic Disruptions:

Certain gut bacteria also play a role in the regulation of circadian rhythms and metabolic processes, both of which are critical for maintaining healthy sleep patterns, as summarized in Table 2. For instance, *Bacteroides*, a signature bacterium found in patients with paradoxical insomnia (P-IN), is positively correlated with poor sleep quality and disrupted circadian rhythms [61,62]. In addition, *Ruminococcaceae* species, including *Ruminococcaceae UCG-002* and *Ruminococcaceae UCG-003*, are associated with chronic insomnia and cardiometabolic diseases, suggesting a link between gut microbiota, sleep disorders, and metabolic health [54,66].

## 4. Sleep-immune-microbiota axis

### 4.1 Introduction to Sleep-Immune-Microbiota Axis

Dynamic interactions between gut microbiota and a host's innate and adaptive immune systems are essential in maintaining intestinal homeostasis and inhibiting inflammation. Gut microbiota metabolizes proteins and complex carbohydrates, synthesizes vitamins, and produces numerous metabolic products that mediate cross-talk between the gut epithelium and immune cells [73]. The composition of the intestinal microbiome plays a pivotal role in maintaining the stability of the intestinal barrier. Dysbiosis contributes to the disruption of this barrier, commonly referred to as "leaky gut" a condition characterized by increased intestinal permeability. This condition facilitates the translocation of bacterial metabolites and endotoxins, such as LPS into the bloodstream, triggering systemic inflammation and facilitating the development of metabolic and autoimmune diseases [74]. Figure 1 illustrates the integrated bidirectional interactions between sleep, the gut microbiota, the immune system, and the endocrine system that frame the conceptual basis of this section.

Comprehending these mechanisms highlights potential therapeutic strategies, such as the use of probiotics, prebiotics, or dietary interventions, to restore gut barrier integrity and mitigate sleep disturbances. This section explores the intricate relationships between sleep, immunity, and microbiota, offering insights into their combined role in maintaining systemic health.

### 4.2 Disbiosis, Inflammation, and Barrier Integrity

Restoring gut microbial balance through fecal microbiota transplantation (FMT) has been shown to reduce LPS levels in the colon, serum, and other tissues, thereby suppressing the TLR4/MyD88/NF-κB signaling pathway and its downstream pro-inflammatory products [75]. This inflammation compromises blood-brain barrier (BBB) integrity, facilitating the translocation of inflammatory mediators and metabolites into the brain. These processes contribute to neuroinflammation, neurodegeneration, and brain aging [76].

Several gut microbiotas, especially *Firmicutes* and *Bacteroidetes*, have demonstrated significant effects on mental health. Dysbiosis involving these groups is associated with mental disorders such as anxiety, depression, and chronic intestinal inflammation [77,78]. An increase in opportunistic pathogens, such as *Aeromonas*, destabilizes intestinal tight junction proteins, allowing microorganisms or microbial components like LPS to enter systemic circulation. This process triggers systemic inflammation, with LPS reaching the brain and binding to Toll-like receptor 4 (TLR4) on microglia, inducing the synthesis and secretion of pro-inflammatory cytokines [13]. Conversely, beneficial bacteria such as *Lactobacillus, Muribaculum*, and *Parasutterella* have been shown to enhance the integrity of both intestinal and brain barriers [45].

### 4.3 Impact of Maternal Sleep Deprivation (MSD) on Gut and Immune Health

Maternal sleep deprivation (MSD) alters gut microbiota and immune responses in offspring. Studies using quantitative real-time polymerase chain reaction (qRT-PCR) and enzyme-linked immunosorbent assay (ELISA) revealed significantly higher expression levels of pro-inflammatory cytokines, such as interleukin 1β (IL-1β) and tumor necrosis factor α (TNF-α), in offspring of MSD-exposed mothers compared to controls. Notably, Ruminococcus_1 and Ruminococcaceae_UCG-005 were positively correlated with these cytokines, suggesting a role in MSD-related neuroinflammation [65].

### 4.4 Sleep, Microbiota, and Immune Dysregulation

Studies on insomnia have demonstrated significant differences in gut microbiota composition between patients and healthy controls. Insomniacs exhibited an increased relative abundance of *Lactobacillus*, *Streptococcus*, and *Lactobacillus crispatus*. These changes were associated with elevated IL-1β levels and reduced TNF-α levels. Specific bacterial shifts, such as increases in *Prevotella* species, were linked to altered immune markers, highlighting the role of microbial metabolites in insomnia pathophysiology [69].

Further research into the interplay between sleep, immune function, and microbiota revealed that immunization with heat-killed *Mycobacterium vaccae* (MV), an environmental bacterium with immunoregulatory properties, mitigates systemic inflammation and behavioral changes induced by sleep disruption. MV immunization prevented alterations in non-REM (NREM) and REM sleep, stress-induced hyperlocomotion, and memory deficits, underscoring the therapeutic potential of microbiota-immune interactions in modulating sleep deprivation [79].

### 4.5 Sleep Disturbance, Immune Activation, and Disease Risk

Sleep disturbances contribute to inflammation-mediated diseases, including depression, through the activation of the innate immune system and an increased risk of infections. Sleep architecture involves dynamic shifts between T helper 1 (Th1)-mediated inflammation during early sleep and T helper 2 (Th2)-mediated responses in late sleep [80]. A study utilizing mass cytometry and single-cell RNA sequencing revealed that sleep deprivation increases T and plasma cell frequencies while upregulating autoimmune-related pathways in CD4+ T and B cells. Sleep deprivation also reduces cytotoxic cell activity, increasing susceptibility to infections and tumor development, while promoting myeloid inflammation and cellular senescence [81].

Severe sleep deprivation in mouse models has been linked to significant inflammation and high mortality. Specifically, increased prostaglandin D2 (PGD2) levels in the brain were shown to drive peripheral immune pathologies, including neutrophil accumulation and cytokine-storm-like syndromes. Disrupting the PGD2/DP1 axis significantly reduced these inflammatory effects, suggesting it as a potential therapeutic target [82].

Obstructive sleep apnea (OSA) disrupts systemic immune function. Single-cell transcriptomics (scRNA-seq) analysis revealed OSA-induced transcriptional changes in peripheral blood mononuclear cells (PBMCs), with severity-dependent alterations in several immune cell lineages. A molecular signature of 32 genes effectively distinguished OSA patients from controls, highlighting deregulation in systemic immunity [83]. Data from the National Health and Nutrition Examination Survey (NHANES) also revealed positive associations between sleep disorders and the systemic immune-inflammation index (SII), with higher SII levels in individuals experiencing sleep problems [84].

Sleep loss induces significant changes in immune cell composition and function, particularly in effector CD4+ T cells and myeloid cells. This is mediated by upregulation of Granulocyte-Macrophage Colony-Stimulating Factor (GM-CSF), a cytokine that drives the IL-23/Th17/GM-CSF feedback mechanism, exacerbating inflammatory responses and autoimmune conditions such as experimental autoimmune uveitis (EAU). Targeting GM-CSF offers a promising therapeutic avenue for managing sleep-related inflammatory diseases [85].

## 5 The Sleep-Endocrine-Microbiota Axis: Interactions and Implications for Health

### 5.1 Sleep and the Endocrine System: An Intimate Relationship

Sleep and the endocrine system share a bidirectional relationship, where hormonal regulation influences sleep quality, and sleep, in turn, modulates endocrine function. Key hormones such as melatonin, cortisol, leptin, ghrelin, and growth hormone (GH) play pivotal roles in this interplay [86–89].

Melatonin, produced by the pineal gland, is a central regulator of the circadian rhythm. Its secretion is stimulated by darkness and inhibited by light, making it essential for synchronizing sleep-wake cycles [86]. Beyond its role in sleep regulation, melatonin exhibits antioxidant and anti-inflammatory properties, which contribute to cellular protection and the mitigation of oxidative stress [90]. Disruptions in melatonin production, such as those caused by exposure to artificial light at night, have been linked to sleep disorders and metabolic dysregulation [91].

Cortisol, a glucocorticoid released by the adrenal glands, follows a diurnal rhythm with peak levels in the early morning and a gradual decline throughout the day. Sleep disturbances, such as insomnia or sleep apnea, can dysregulate cortisol secretion, leading to hyperactivation of the hypothalamic-pituitary-adrenal (HPA) axis. This dysregulation is associated with increased stress, anxiety, and a higher risk of metabolic and cardiovascular diseases [92]. For instance, sleep deprivation has been shown to elevate nighttime cortisol levels, exacerbating stress-related disorders [93].

Studies have shown that changes in the ghrelin/leptin ratio are significantly correlated with alterations in subjective hunger during chronic circadian disruption and sleep restriction [94]. Growth hormone (GH) is primarily secreted during slow-wave sleep (SWS), and its release is regulated by growth hormone-releasing hormone (GHRH) and somatostatin. GH stimulates the production of insulin-like growth factor 1 (IGF-1), which plays a key role in tissue growth, neuroprotection, and metabolic regulation. Sleep deprivation significantly reduces GH and IGF-1 levels, affecting metabolic homeostasis and cognitive function [89].

**5.2 The Role of the Endocrine System in the Microbiota-Sleep Connection**

The gut microbiota, through its influence on the intestinal environment and systemic pathways, is recognized as a functional endocrine organ. It interacts with various hormones, including estrogen, androgens, and insulin, playing a critical role in endocrine regulation [95]. Recent studies have shown that gut dysbiosis can activate the hypothalamic-pituitary-adrenal (HPA) axis, leading to a hormonal imbalance that impacts sleep patterns. This activation of the HPA axis can result in elevated cortisol levels, a stress hormone that, when excessive, can disrupt sleep architecture and contribute to sleep disorders [96].

Additionally, gut microbiota influences the production of serotonin, a neurotransmitter that serves as a precursor to melatonin, the primary hormone regulating the sleep-wake cycle. An imbalance in the microbiota can alter serotonin [97].

The relationship between the microbiota and the endocrine system is also evident in glucose metabolism regulation. A study observed that reduced REM sleep duration is associated with an unfavorable glycemic profile and alterations in microbiota composition, suggesting an interaction between sleep, microbiota, and endocrine regulation of energy metabolism [98].

Moreover, interventions that modulate the gut microbiota, such as the use of probiotics and prebiotics, have shown promising effects in improving sleep disorders. These interventions may influence the endocrine system by restoring microbiota balance, normalizing the production of hormones and neurotransmitters involved in sleep regulation [99].

In summary, the endocrine system acts as a key mediator in the bidirectional connection between gut microbiota and sleep. Alterations in the microbiota can trigger endocrine responses that affect sleep quality, while sleep disturbances can influence microbial composition and endocrine function. Understanding this interaction is essential for developing therapeutic strategies aimed at improving sleep health through microbiota and endocrine system modulation.

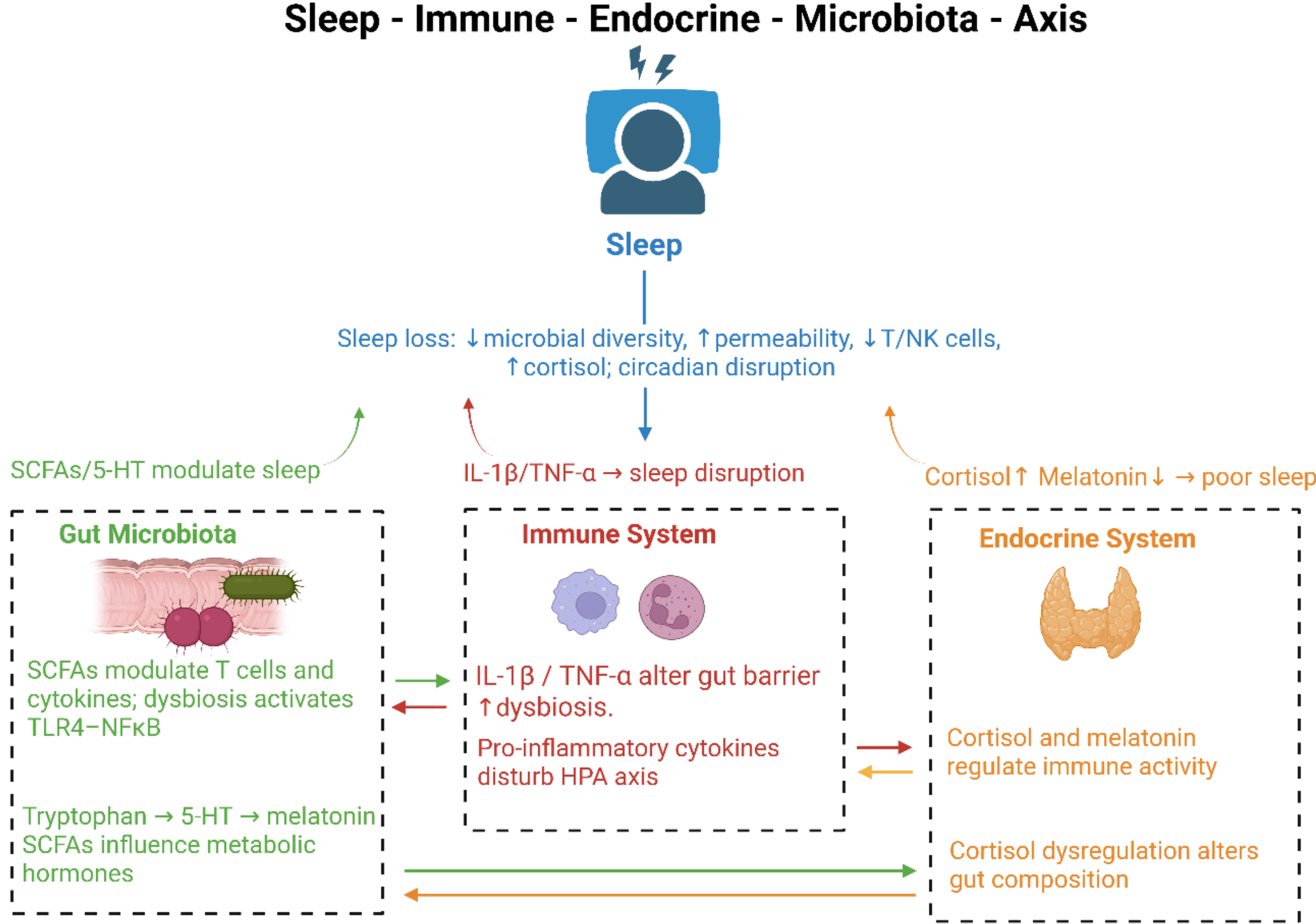

**Figure 1. Integrated bidirectional interactions between sleep, gut microbiota, the immune system, and the endocrine system.**

Sleep loss reduces microbial diversity, increases intestinal permeability, lowers T/NK cell activity, elevates cortisol, and disrupts circadian rhythms. Gut dysbiosis alters SCFA and serotonin production, activates TLR4–NFκB signaling, and promotes inflammation. Immune activation (IL-1β, TNF-α) disrupts sleep architecture, damages the gut barrier, and stimulates HPA axis activity. Endocrine dysregulation is characterized by increased cortisol and reduced melatonin. This further impairs sleep and modifies gut microbial composition. Together, these systems form a tightly interconnected axis in which disturbances in one component propagate through the others.

## 6.0 Emerging Therapies Targeting the Sleep-Microbiota Axis

The sleep-microbiota axis has emerged as a promising frontier in the development of novel therapeutic strategies. The intricate bidirectional relationship between sleep, gut microbiota, and the immune system offers multiple intervention points to enhance sleep quality and reduce systemic inflammation. Probiotic interventions, particularly with strains such as *Lactobacillus* and *Bifidobacterium*, have demonstrated potential in improving sleep by modulating the gut-brain axis, altering neurotransmitter synthesis, and reducing inflammatory markers associated with sleep disorders [100]. In parallel, emerging nutraceutical approaches, such as compositions including β-glucan, prebiotics, and the herbal extract silymarin, have shown promise. A 90-day pilot study demonstrated that these formulations improved sleep quality, mood, and life quality while reducing inflammatory markers and enhancing metabolic health, suggesting their potential as integrative therapies targeting the sleep-microbiota axis [101]. As illustrated in Figure 2, therapeutic interventions targeting the sleep-microbiota axis encompass probiotics, fecal microbiota transplantation, nutraceuticals, neuropeptides, chrononutrition, and acupuncture, all of which act on microbiota composition, circadian regulation, neurochemical balance, and microbe-host communication.

Emerging therapeutic approaches continue to explore the potential of fecal microbiota transplantation (FMT) and its advanced variant, washed microbiota transplantation (WMT), as innovative interventions to restore gut integrity and mitigate neuroinflammation. Notably, WMT has shown promising results in improving sleep quality among patients with inflammatory bowel disease (IBD) [102]. Moreover, a recent study demonstrated that WMT significantly enhances sleep quality and life quality in patients with sleep disorders by regulating gut microbiota, with improved outcomes observed following multiple treatment courses. These findings highlight WMT's safety and efficacy, further supporting its role as a novel therapeutic option for targeting the sleep-microbiota axis [103].

Furthermore, advancements in chronobiology have revealed that gut microbiota exhibits a circadian rhythm closely synchronized with host sleep-wake cycles. Dysbiosis disrupts this rhythm, negatively impacting sleep quality. Therapies targeting the restoration of microbiota circadian patterns are gaining attention as a strategy for managing sleep disorders [104]. For instance, chrononutrition, or the timing of food intake in alignment with circadian rhythms, has shown potential to enhance microbial rhythmicity and improve sleep outcomes. Chrononutrition, which includes practices such as diurnal fasting, meal timing, and avoiding late eating, has been linked to improvements in sleep quality, particularly through its influence on metabolic regulation and circadian alignment [105].

Microbial metabolites such as short-chain fatty acids (SCFAs) are increasingly being investigated for their roles in sleep regulation, particularly in the context of insomnia.

SCFAs, including acetate, butyrate, and propionate, are key byproducts of fiber fermentation in the gut and influence gut-brain communication pathways associated with sleep continuity. Evidence from studies in older adults with insomnia symptoms suggests that higher concentrations of SCFAs are linked to poorer sleep efficiency and longer sleep onset latency, particularly in individuals with the short sleep duration phenotype, which is considered a more biologically severe form of insomnia [106]. These findings highlight the potential of SCFAs not only as biomarkers of sleep disorders but also as attractive targets for future therapeutic interventions.

Additionally, the use of neuropeptides, such as orexins, vasoactive intestinal peptide (VIP) and neuropeptide Y represents another avenue for therapeutic intervention. These molecules modulate both sleep cycles and gut microbiota interactions, offering a potential dual-target strategy for improving sleep and gut health is essential for wakefulness maintenance, while melanin-concentrating hormone and galanin promote REM sleep [107]. In parallel, neuropeptide S (NPS) has been shown to alleviate anxiety-like behavior and sleep disturbances caused by paradoxical sleep deprivation (PSD). NPS modulates wakefulness, suppresses paradoxical sleep, and alters EEG theta activity. By activating NPSR receptors in the amygdala, NPS counteracts the effects of PSD without triggering rebound sleep, which further underscores its potential as a therapeutic option for anxiety-related sleep disorders. [108].

Complementary therapies, such as acupuncture, have demonstrated potential as nonpharmacological interventions for insomnia by modulating both neurotransmitter levels and gut microbiota composition. In PCPA (p-chlorophenylalanine)-induced insomnia models, acupuncture was shown to reduce serum levels of dopamine, 5-hydroxytryptamine, and norepinephrine, while increasing melatonin levels in the pineal gland. Furthermore, 16S rRNA sequencing revealed that both acupuncture and hypnotic drugs produced similar improvements in gut microbiota composition, suggesting shared mechanisms of action. Notably, acupuncture networks exhibited greater microbial community stability and fewer side effects compared to pharmacological interventions, highlighting its potential as a safer alternative for targeting the sleep-microbiota axis [109].

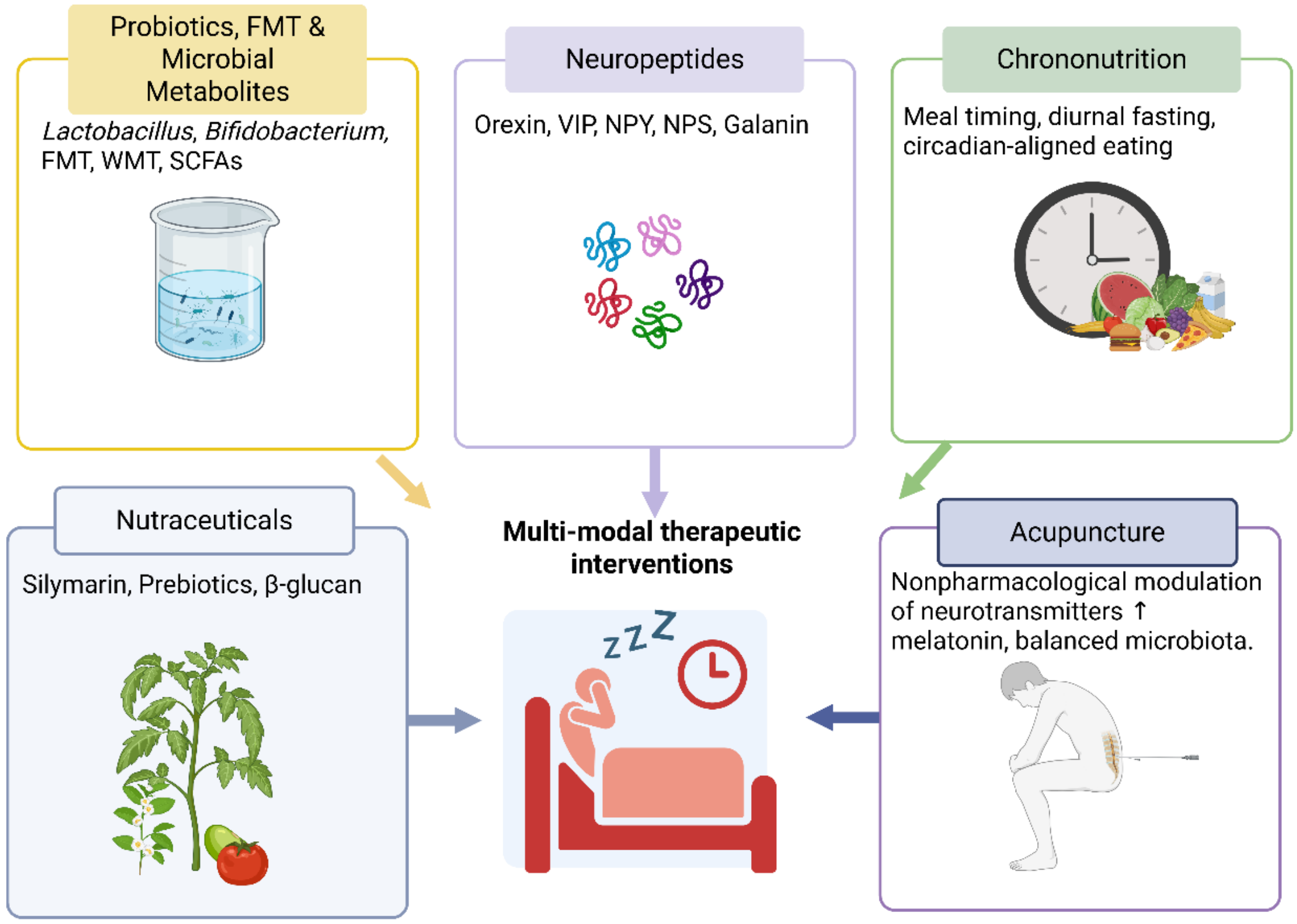

**Figure 2. Emerging therapeutic strategies targeting the sleep–microbiota axis.**

Multiple approaches aim to restore sleep quality by modulating gut microbial composition, neuroendocrine signaling, and circadian alignment. Probiotics, fecal and washed microbiota transplantation (FMT/WMT), and microbial metabolites such as SCFAs support microbial balance and gut–brain communication. Nutraceuticals, including silymarin, prebiotics, and β-glucan, enhance metabolic and anti-inflammatory pathways relevant to sleep regulation. Neuropeptides such as orexin, VIP, NPY, NPS, and galanin offer additional targets for modulating arousal and sleep architecture. Chrononutrition strategies diurnal fasting, meal timing, and circadian-aligned eating reinforce microbial and metabolic rhythms. Acupuncture serves as a nonpharmacological therapy that influences neurotransmitters, increases melatonin, and promotes a healthier microbiota. Together, these modalities illustrate a multi-modal therapeutic framework for improving sleep outcomes.

### 7.0 Limitations of Current Literature

Despite major advances in understanding the sleep microbiota axis, several limitations constrain the current body of evidence.
First, most studies rely on small sample sizes, reducing statistical power and generalizability. For example, clinical work examining fecal metabolites and sleep quality uses modest cohorts that limit robust subgroup analyses and mechanistic inference [22].

Second, much of the mechanistic evidence still comes from animal models, particularly rodent studies on circadian disruption, sleep loss, and neuroinflammation. While informative, these models cannot fully capture the complexity of human microbiota, or the variability in lifestyle, diet, and circadian behavior seen in real populations [45].

Third, methodological heterogeneity across microbiome studies remains a major limitation. Differences in sequencing platforms, microbial classification pipelines, and metabolite profiling approaches create inconsistencies across reports, complicating direct comparisons between findings.
Studies evaluating interventions such as nutraceutical compositions or acupuncture further illustrate variability in outcome measures, microbial endpoints, and inflammatory biomarkers, making it difficult to establish unified mechanistic pathways [101,109].

Although endocrine pathways are clearly affected by sleep loss, this component of the axis remains understudied. Current evidence shows that sleep deprivation and circadian disruption alter cortisol rhythms and HPA-axis activity, contributing to metabolic and inflammatory imbalance [87]. However, most studies assess cortisol or sleep outcomes in isolation, without simultaneous measurement of immune or microbiota variables [92].

As a result, very few human studies integrate endocrine, immune, and microbial markers within a unified experimental design. Finally, therapeutic approaches including chrononutrition, nutraceutical formulations, and microbiota transplantation have shown promising but preliminary results with significant variability in dosage, duration, and treatment adherence [75,101,105] These limitations highlight the need for larger, standardized, multimodal studies to fully understand the bidirectional interactions among sleep, microbiota, endocrine signaling, and immune networks.

## 8.0 Future Directions for Research

Future research should integrate microbiota, neurochemical signaling, immune activity, and endocrine rhythms to define causal mechanisms of the sleep microbiota axis. Sleep-loss–driven IL-23/Th17/GM-CSF inflammation [85] together with dysbiosis induced TLR4/NF-κB activation [75] highlight the need for interventions targeting barrier integrity and systemic inflammation. Mechanistic studies must clarify how sleep-regulating neurotransmitters described in sleep-wake neurochemistry are shaped by GABA-producing bacteria [29], SCFAs, and disrupted serotonin vitamin B6 metabolism in antibiotic depleted models [110].

Therapeutically, both WMT improving sleep in humans [102] and acupuncture modulating microbiota and monoamines in insomnia models [109] suggest promising microbiota directed interventions. Future work should also evaluate chrononutrition [105] and metabolic circadian alignment, particularly in disorders affecting leptin/ghrelin oscillations. Finally, deeper characterization of GMBA immune neural interactions [20] is essential for developing integrated neuroimmune microbial therapies for sleep disorders.

**Conclusions:**

The interplay between the gut microbiota and sleep regulation represents a fascinating and rapidly evolving area of research. While substantial progress has been made, critical gaps in knowledge remain. For instance, the mechanisms by which specific microbial metabolites influence sleep architecture are still not fully understood. Moreover, the variability in individual microbiota compositions presents challenges in developing universally effective therapies.

Evidence suggests that targeting microbiota may offer novel therapeutic avenues for sleep disorders, particularly those with an inflammatory or stress-related component. Future therapies could integrate microbiota modulation with established clinical protocols, such as cognitive-behavioral therapy for insomnia or pharmacological interventions. Additionally, emerging diagnostic tools, like microbiota-based biomarkers, hold promise for identifying individuals at risk of sleep disorders or monitoring treatment efficacy.

Future investigations should prioritize integrative approaches that combine microbiota-targeted therapies with lifestyle modifications, such as dietary adjustments, sleep hygiene, and stress management. By addressing these aspects holistically, researchers and clinicians can maximize the potential of these therapies to improve sleep quality and overall health outcomes.

**Acknowledgements**

Figures in this manuscript were created with BioRender.com.